\documentclass[12pt]{iopart}

\pdfoutput=1

\usepackage{iopams}  

\usepackage{amstext}
\usepackage{amsfonts}
\usepackage{graphicx}
\usepackage{subfig}
\usepackage{mathbbol}
\usepackage{color}

\usepackage[utf8]{inputenc}
\usepackage[greek,english]{babel}
\usepackage{lmodern}

\def\Tr{\mbox{Tr}}

\begin{document}

\title[]{Maximal energy extraction via quantum measurement}
\author{Andrea Solfanelli}
\address{Department of Physics and Astronomy, University of Florence, Via Sansone 1, I-50019, Sesto Fiorentino (FI), Italy.}
\address{INFN Sezione di Firenze, via G. Sansone 1, I-50019, Sesto Fiorentino (FI), Italy.}

\author{Lorenzo Buffoni}
\address{Department of Physics and Astronomy, University of Florence, Via Sansone 1, I-50019, Sesto Fiorentino (FI), Italy.}
\address{Department of Information Engineering, University of Florence, Via di S. Marta 3, 50139, Firenze (FI), Italy.}

\author{Alessandro Cuccoli}

\address{Department of Physics and Astronomy, University of Florence, Via Sansone 1, I-50019, Sesto Fiorentino (FI), Italy.}
\address{INFN Sezione di Firenze, via G. Sansone 1, I-50019, Sesto Fiorentino (FI), Italy.}

\author{Michele Campisi}

\address{Department of Physics and Astronomy, University of Florence, Via Sansone 1, I-50019, Sesto Fiorentino (FI), Italy.}
\address{
INFN Sezione di Firenze, via G. Sansone 1, I-50019, Sesto Fiorentino (FI), Italy.}

\begin{abstract}
We study the maximal amount of energy that can be extracted from a finite quantum system by means of projective measurements. For this quantity we coin the expression ``metrotropy" $\mathcal{M}$, in analogy with ``ergotropy" $\mathcal{W}$, which is the maximal amount of energy that can be extracted by means of unitary operations. The study is restricted to the case when the system is initially in a stationary state, and therefore the ergotropy is achieved by means of a permutation of the energy eigenstates. We show that i) the metrotropy is achieved by means of an even combination of the identity and an involution permutation; ii) it is $\mathcal{M}\leq\mathcal{W}/2$, with the bound being saturated when the permutation that achieves the ergotropy is an involution.
\end{abstract}

\maketitle

\section{Introduction}
Since the seminal works of Hatsopoulos \cite{Hatsopoulos76FP6b} and Allahverdyan \emph{et al.} \cite{Allahverdyan04EPL67} the concept of ergotropy (namely the maximal amount of energy that can be extracted from a quantum system by means of unitary operations), has become central in the field of quantum thermodynamics. Motivated by previous works that study the impact of quantum measurements on thermodynamic processes \cite{Campisi10PRL105,Campisi11PRE83,Watanabe14PRE89,Talkner16PRE93,Schulman06PRL97}
and consider the measurement process itself as a thermodynamic resource that may be employed to fuel quantum heat engines \cite{Watanabe17PRL118,Elouard17NPJQI3,Elouard18PRL120,Buffoni19PRL122}, here we consider the question of what is the maximal amount of energy that can be extracted from a  quantum system by means of projective measurements. This is of particular interest for the development of novel cooling mechanisms that exploit genuinely quantum mechanical effects. For this quantity we coin the expression ``metrotropy'' (from \textgreek{m\'etron}: measure, and \textgreek{trop\'h}: change) in analogy with Allahverdyan's \emph{et al.} ``ergotropy'' (from \textgreek{\'ergon}: work, and \textgreek{trop\'h}: change \cite{Allahverdyan04EPL67}).

We focus on the case when the quantum system is initially in a stationary state, namely its density matrix $\rho$ commutes with the system Hamiltonin $H$. In this case the ergotropy, $\mathcal W$, is achieved by a unitary transformation that realises the permutation $\boldsymbol{\sigma}_\mathcal{W}$ of the energy eigenstates that orders them according to their population (highest populations to lowest energies) \cite{Allahverdyan04EPL67}. Here we show that the metrotropy, $\mathcal M$, is achieved by a projection channel realising an even linear combination of the identity and an involution (an involution is a permutation that does not contain any permutation cycles of length larger than 2). We show that the metrotropy generally  reads
\begin{eqnarray}
\mathcal{M}=(E-v_0)/2 \leq \mathcal W/2
\end{eqnarray}
where $E = \Tr H\rho$ is the system initial energy and $v_0$ is the smallest energy reachable by means of an involution.  When the permutation that achieves the ergotropy is itself an involution then the metrotropy is half the ergotropy.

As we shall see below, the presented result is a consequence of the strong restrictions imposed by the requirement that a doubly-stochastic matrix entering the expression of energy extraction, be unitary stochastic (in short unistochastic). 
The study of the geometry and topology of the subset of the set of  bistochastic matrices which are unistochastic is a topic of interest within the field of quantum information theory \cite{BengtssonBook} and as well in the foundations of quantum theory \cite{Lande57PR108,Rovelli96IJTP35}. It is sometimes stated or implied in the literature that any doubly stochastic matrix is unistochastic \cite{Lande57PR108,Rovelli96IJTP35,Allahverdyan08PRE77}. That is in fact only true for $2\times 2$ matrices but not generally true for $N \times N$ matrices, with $N>2$ \cite{BengtssonArXiv0403088}. For example, the following $3\times 3$ bistochastic matrix
\begin{eqnarray}
\frac{1}{2}\left(\begin{array}{ccc}1 & 1 & 0 \\0 & 1 & 1 \\1 & 0 & 1 \end{array}\right)
\end{eqnarray}is not unistochastic \cite{Zyczkowski03JPA36}.

\section{Ergotropy}
We consider a quantum system with Hamiltonian $H$ that is initially in state $\rho$, and assume the state is stationary $[\rho,H]=0$.  The system is acted upon by means of external time-dependent fields within some time lapse $\tau$. Its evolution is accordingly governed by a unitary map $U$:
\begin{equation}
\rho' = U\rho U^\dagger
\end{equation}
We are interested in the average work done on the system
\begin{equation}
W = E'-E = \Tr H \rho' - \Tr H \rho
\end{equation}
Let 
\begin{eqnarray}
H= \sum_k E_k |k\rangle \langle k |  \\
\rho = \sum_ k r_k  |k\rangle \langle k |
\end{eqnarray}
then,
\begin{eqnarray}
W = \sum_k E_k \langle k| U \rho U^\dagger |k\rangle - \sum E_k  r_k  = 
\sum_{k,l }E_k P_{kl}r_l - \sum E_k  r_k 
\end{eqnarray}
where 
\begin{eqnarray}
P_{kl}= |\langle k| U | l \rangle|^2 \, .
\end{eqnarray}
We are interested in extracting the maximal amount of energy, namely we want to maximise $-W$, or, equivalently, minimise the system final energy expectation $E' =\sum_{kl} E_k P_{kl}r_l$ over all unitaries $U$. In the following we shall use, for simplicity the matrix notation $E' = \mathbf{E}^T \cdot\mathbf{P} \cdot \mathbf{r}$. The minimum of the expression $ \mathbf{E}^T \cdot\mathbf{P} \cdot \mathbf{r} $ is reached when $\mathbf{P}$ is the permutation $\boldsymbol{\sigma}_\mathcal{W}$ that maps the largest among the populations $r_l$ to the smallest among the energies $E_k$, the second largest population to the second smallest energy, and so on \cite{Allahverdyan04EPL67}. 

As an example consider a 3-state system  with eigenenergies $\mathbf{E} = (0,\varepsilon, 2\varepsilon)^T$ and populations $ \mathbf{r}= (0.2, 0.5, 0.3)^T$. The smallest final energy $E' = 0.5 \cdot 0 +0.3 \cdot \varepsilon + 0.2  \cdot2\varepsilon=0.7\varepsilon$ is reached when $\mathbf{P}$ is the permutation matrix
\begin{eqnarray}
\mathbf{\boldsymbol{\sigma}}_\mathcal{W}=\left(\begin{array}{ccc}0 & 0 & 1 \\1 & 0 & 0 \\0 & 1 & 0\end{array}\right)
\label{eq:exampleSigma}
\end{eqnarray}
The permutation $\boldsymbol{\sigma}_\mathcal{W}$ is realisable with unitary operations of the form  
\begin{eqnarray}
U_\mathcal{W}=\left(\begin{array}{ccc}0 & 0 & e^{i\alpha} \\e^{i\beta} & 0 & 0 \\0 & e^{i\gamma} & 0\end{array}\right)
\end{eqnarray}
\section{Metrotropy}
We consider a quantum system with Hamiltonian $H$ that is initially in state $\rho$, and assume the state is stationary $[\rho,H]=0$. The system interacts with a measurement apparatus, having the effect of projecting the state onto some projection basis:
\begin{equation}
\rho' = \sum_k \pi_k \rho \pi_k
\end{equation}
where the $\pi_k$'s form a complete set of projectors: $\sum_k \pi_k = \mathbb{1}$, $\pi_k\pi_l= \delta_{kl} \pi_l$. We shall assume that the projectors $\pi_k$ are of rank one, namely one can write them as
\begin{equation}
\pi_k = |\psi_k\rangle \langle \psi_k | \, .
\end{equation}
The post measurement energy $E'$ then reads
\begin{eqnarray}
E' &= \sum_{lk} E_l \langle l|\psi_k\rangle \langle \psi_k | \rho |\psi_k\rangle \langle \psi_k  | l\rangle  \\
	&= \sum_{lkn} E_l \langle l|\psi_k\rangle \langle \psi_k | n\rangle \langle n |\psi_k\rangle \langle \psi_k  | l\rangle r_n \\
	&= \sum_{lkn} E_l | \langle l|\psi_k \rangle |^2  | \langle n |\psi_k\rangle |^2  r_n \\
	&= \mathbf{E}^T \cdot\mathbf{P}^T \cdot\mathbf{P} \cdot \mathbf{r}
\end{eqnarray}
where $P_{nk}= |\langle n | \psi_k \rangle |^2$ denotes the probability to find the system in state $|\psi_k\rangle$ after the measurement has occurred, provided it was prepared in state $|n \rangle$. Since the states $|\psi_k\rangle$ form a basis for the system's Hilbert space, there exist some unitary  $U$, such that $|\psi_k\rangle = U |k\rangle$, then $P_{nk}=|\langle n |U|k\rangle |^2$.

Matrices of the form $P_{kl}=|\langle k|U|l\rangle |^2$ with $U$ a unitary operator, are called uni-stochastic matrices \cite{Marshall11book}. A bistochastic matrix $B_{kl}$ is a square matrix satisfying the conditions; i) $B_{kl}\geq 0$ for all $k,l$; ii) $\sum_k B_{kl} = \sum_l B_{kl} =1$. It is easy to see that any uni-stochastic matrix is bistochastic. The converse is not generally true unless the matrix is $2\times 2$ \cite{BengtssonArXiv0403088}. It is also easy to see that the product of two bistochastic matrices is itself bi-stochastic. 

We are interested in finding the minimum of $E' = \mathbf{E}^T \cdot\mathbf{P}^T \cdot\mathbf{P} \cdot \mathbf{r}$ over all unistochastic matrices $\mathbf{P}$, that is over all unitaries via the relation $P_{kl}=|\langle k|U|l\rangle |^2$. We first note that, once the ergotropy permutation $\boldsymbol{\sigma}_{\mathcal{W}}$ is known, it is generally impossible to express it as the ``square'' $\mathbf{P}^T \cdot\mathbf{P}$ of some bistochastic matrix $\mathbf{P}$. 
This can be seen by considering the Birkhoff theorem according to which any bistochastic matrix can be written as a convex combination of permutations:
\begin{equation}
\mathbf{P} = \sum_i \lambda_i \boldsymbol{\sigma}_i \, ,
\label{eq:Birkoff}
\end{equation}
where $\boldsymbol{\sigma}_i$ are permutation matrices and $\lambda_i \geq 0, \sum_i \lambda_i=1$. Then $\mathbf{P}^T\cdot \mathbf{P}=\sum_{ij}\lambda_i\lambda_j \boldsymbol{\sigma}_i^T \boldsymbol{\sigma}_j$. Note that the product $\boldsymbol{\sigma}_i^T \boldsymbol{\sigma}_j$ of two permutations is itself a permutation, note also that $\sum_{i,j}\lambda_i\lambda_j=1, \lambda_i\lambda_j\geq0$. That is the double sum $\sum_{ij}\lambda_i\lambda_j \boldsymbol{\sigma}_i^T \boldsymbol{\sigma}_j$ is itself a convex combination of permuatations (recall that $\mathbf{P}^T\cdot \mathbf{P}$ is itself doubly stochastic). In order for this sum to be itself a single permutation one needs that only one among the coefficients $\lambda_i\lambda_j$ equals $1$ and all the other are null. This means that there is one label $i^*$ such that $\lambda_i^*=1$, and $\lambda_{i^*\neq i}=0$. Then the double sum reduces to $\boldsymbol{\sigma}_{i^*}^T\boldsymbol{\sigma}_{i^*}=\mathbb{1}$, because the transpose of a permutation is its inverse. That is, unless $\boldsymbol{\sigma}_\mathcal{W}=\mathbb{1}$,  $\boldsymbol{\sigma}_\mathcal{W}$ cannot be written in the form  $\mathbf{P}^T\cdot \mathbf{P}$. This implies that the metrotropy generally does not coincide with the ergotropy. 

Our first step towards the proof of our main result consists of finding the minimum of $E' = \mathbf{E}^T \cdot\mathbf{P}^T \cdot\mathbf{P} \cdot \mathbf{r}$ over all bistochastic matrices $\mathbf{P}$. The theorem would then follow by further imposing a necessary condition for  $\mathbf{P}$ to be unistochastic. 
Using Eq. (\ref{eq:Birkoff}) we have
\begin{eqnarray}
E' = \sum_{i j } \lambda_i \lambda_j  \mathbf{E}^T \cdot \boldsymbol{\sigma}_i^T \cdot \boldsymbol{\sigma}_j \cdot \mathbf{r} =  \sum_{i j } \lambda_i \lambda_j u_{ij}\, ,
\end{eqnarray}
where
the $u_{ij}$'s form a symmetric $N! \times N!$ real matrix
\begin{eqnarray}
u_{ij}=  \mathbf{E}^T \cdot \frac{\boldsymbol{\sigma}_i^T \boldsymbol{\sigma}_j + \boldsymbol{\sigma}_j^T \boldsymbol{\sigma}_i}{2}\cdot  \mathbf{r}
\end{eqnarray}
We note that $E$, the initial energy, is associated to the choice $\boldsymbol{\sigma}_i=\boldsymbol{\sigma}_j=\mathbb{1}$. 
If $E$ is the smallest among the $u_{ij}$ then the metrotropy is trivially null. 
Let $u_{0}$ be the smallest among the $u_{ij}$'s, $u_{1}$ be the second smallest and so on. Let's say $E$ is the $(k+1)^{th}$ in the list.
For simplicity we shall assume for now that $k=2$, that is there is only $u_1$ between $u_0$  and $E$ (the treatment of the more general case proceeds straightforwardly), that is we assume the ordering
\begin{eqnarray}
u_{0}\leq u_1 \leq u_{2}=E \leq \dots 
\label{eq:barU}
\end{eqnarray}
For sake of simplicity we assign the labels $1,2$ to the couple of permutations associated to $u_0$: $u_0=u_{1,2}$.
This can always be done because if $i^*,j^*$ are  such that $u_{i^*,j^*}=u_0$, since $\boldsymbol{\sigma}_{i^*}^T \boldsymbol{\sigma}_{j^*}$ is a permutation, there is a $k^*$ such that $\boldsymbol{\sigma}_{i^*}^T \boldsymbol{\sigma}_{j^*}=\boldsymbol{\sigma}_{k^*}=\boldsymbol{\sigma}_1^T \boldsymbol{\sigma}_{k^*}$, where $\boldsymbol{\sigma}_1$ is the identity. Hence $\boldsymbol{\sigma}_{i^*}^T \boldsymbol{\sigma}_{j^*} + \boldsymbol{\sigma}_{j^*}^T \boldsymbol{\sigma}_{i^*} =\boldsymbol{\sigma}_1^T \boldsymbol{\sigma}_{k^*}+  \boldsymbol{\sigma}_{k^*}^T \boldsymbol{\sigma}_1$, and $u_{i^*,j^*}=u_0= u_{k^*,1}$, that is we can replace $i^*,j^*$ with $k^*,1$. Similarly we assign the labels $1,3$ to the couple of permutations associated to $u_1$: $u_1=u_{1,3}$. 

We have
\begin{eqnarray}
E' &=  \sum_{i\neq j}   \lambda_i \lambda_j u_{ij} + \sum_i \lambda_i^2 E \\
&= \sum_{i\neq j}  u_{ij} \lambda_i \lambda_j + (1-\sum_{i\neq j}  \lambda_i \lambda_j) E \\
&= 2 \lambda_2 \lambda_1 u_{0}  + 2 \lambda_1 \lambda_3 u_1 + \sum'_{i\neq j}   \lambda_i \lambda_j u_{ij} + E -(2\lambda_2 \lambda_1 +2\lambda_1\lambda_3+ \sum'_{i\neq j}  \lambda_i \lambda_j) E \nonumber \\
&= 2 \lambda_2 \lambda_1 (u_{0}-E) + 2 \lambda_1 \lambda_3 (u_{1}-E)  + E + \sum'_{i\neq j}   \lambda_i \lambda_j (u_{ij}-E) \nonumber \\
& \geq 2 \lambda_2 \lambda_1 (u_{0}-E) + 2 \lambda_1 \lambda_3 (u_{1}-E)  + E \\
& \geq 2 \lambda_1(\lambda_2+\lambda_3) (u_{0}-E)  + E \label{eq:Efbound-general} \\
& \geq  (u_{0}+E)/2
\end{eqnarray}
where $\sum'$ is restricted to $(i,j) \neq (1,2),(2,1),(1,3),(3,1)$. The last step follows from the inequality
$\lambda_1(\lambda_2+\lambda_3) \leq 1/4\, $. The argument can be repeated similarly for any $k$, by grouping in the primed sum all the terms such that $u_{ij} > E$.
The bound can be saturated with the choice $\lambda_1=\lambda_2=1/2$ and $\lambda_i=0$ for $i\neq 1,2$. Therefore $(u_{0}+ E)/2$ is the minimum of $E'$ over all possible doubly stochastic $\mathbf{P}$'s and it
is achieved for $\mathbf{P}=(\boldsymbol{\sigma}_1+\boldsymbol{\sigma}_2)/2= (\mathbb{1}+\boldsymbol{\sigma}_2)/2$.

The question now is whether $\mathbf{P}$ is generally unistochastic, and if not what is the minimum value of $E'$ reachable within the subset of bistochastic matrices that are unistochastic.
To answer this question we employ a theorem presented in Ref. \cite{Yik-Hoi91LINALGAPP150} according to which any convex combination of permutation matrices which is unistochastic, is such that it involves only permutations that are pairwise complementary.
Two $N \times N$ matrices $\mathbf{A}$ and $\mathbf{B}$ are said to be complementary if, for any $1\leq i,j,h,k\leq N$, $A_{ij}=A_{hk}=B_{ik}=1$ implies $B_{hj}=1$ \footnote{We warn the reader that  Ref. \cite{Yik-Hoi91LINALGAPP150} uses the therm ``orthostochastic'' to designate what here we refer to as ``unistochastic''. See also \cite{Chterental08LAA428} for the same warning.}. 
It is crucial now to note that any permutation that is complementary to the identity is symmetric \cite{Yik-Hoi91LINALGAPP150}. Hence if $\boldsymbol{\sigma}_2$ is not symmetric, the combination $(\mathbb{1}+\boldsymbol{\sigma}_2)/2$ is not unistocahstic.  
We note also that two symmetric permutations are complementary if and only if they commute with each other \cite{Yik-Hoi91LINALGAPP150}. 

In the light of the mentioned theorem, we add the constraint that the Birkhoff expansion contains only pairwise complementary permutations. Now, looking at Eq. (\ref{eq:Efbound-general}) we see that, if the identity is not included in the sum, i.e., if $\lambda_1=0$, then $E'\geq E$. Hence the expansion must include the identity in order to be able to reach the minimum of $E'$. 
Accordingly we repeat the argument above, but restricting now the expansion to contain only permutations belonging to the set $\mathcal I_N$ of symmetric permutations, with the further constraint that all permutations of the expansion should commute with each other. Thus, we get
\begin{eqnarray}
E' \geq  (v_{0}+E)/2
\end{eqnarray}
where $v_0$ is the smallest among the 
\begin{eqnarray}
v_{ij}  \doteq  \mathbf{E}^T \cdot \boldsymbol{\sigma}_i \cdot \boldsymbol{\sigma}_j \cdot \mathbf{r}, \qquad \boldsymbol{\sigma}_i,\boldsymbol{\sigma}_j \in \mathcal I_N, \, [\boldsymbol{\sigma}_i,\boldsymbol{\sigma}_j]=0
\end{eqnarray}
where $\boldsymbol{\sigma}_i,\boldsymbol{\sigma}_j $ are symmetric commuting permutations. 
We note that the product of two commuting symmetric permutations is itself a symmetric permutation:
$(\boldsymbol{\sigma}_i\boldsymbol{\sigma}_j)^T=\boldsymbol{\sigma}_j^T\boldsymbol{\sigma}_i^T=\boldsymbol{\sigma}_j\boldsymbol{\sigma}_i=\boldsymbol{\sigma}_i\boldsymbol{\sigma}_j$. Accordingly, $v_0$  is the minimum of $ \mathbf{E}^T \cdot \boldsymbol{\sigma} \cdot \mathbf{r}$ over the set, $\mathcal S_N$, of symmetric permutations that are product of two commuting symmetric permutations. But $\mathcal S_N$ coincides with the set $\mathcal I_N$ of symmetric permutations itself.  In fact on one hand it is trivially $\mathcal S_N \subseteq \mathcal I_N$,  and on the other hand, each element in $\mathcal I_N$ can be expressed as the product of itself and the identity (which is symmetric and commutes with every other permutation), hence it belongs to $\mathcal S_N$, implying $\mathcal I_N \subseteq \mathcal S_N$, hence $\mathcal I_N = \mathcal S_N$. Summing up:
\begin{eqnarray}
v_0 = \min_{\boldsymbol{\sigma} \in \mathcal I_N}   \mathbf{E}^T \cdot \boldsymbol{\sigma} \cdot \mathbf{r} \, .
\end{eqnarray}
The bound is saturated by
\begin{equation}
\mathbf{P}=(\mathbb{1}+\boldsymbol{\sigma}_{\mathcal{M}})/2
\end{equation}
with 
\begin{equation}
\boldsymbol{\sigma}_{\mathcal{M}} \doteq \arg\min_{\boldsymbol{\sigma} \in \mathcal I_N}  \mathbf{E}^T \cdot \boldsymbol{\sigma} \cdot \mathbf{r} \, .
\end{equation}
We recall that the condition that a convex combination of permutations contains only pairwise complementary permutations is necessary for the combination to be unistochastic, while sufficiency has been proved, to the best of our knowledge, for $N \leq 15$ only \cite{Yik-Hoi91LINALGAPP150}. Hence it remains to show that our special convex combination of two complementary permutations $\mathbf{P}=(\mathbb{1}+\boldsymbol{\sigma}_\mathcal{M})/2$, with symmetric $\boldsymbol{\sigma}_\mathcal{M}$ is unistochastic, for any $N$. 
To show that, we provide the explicit expression of a unitary operator $U$ such that $P_{ij}= |\langle i | U | j \rangle|^2$. 
We recall that each and all permutations that are represented by symmetric matrices are involution permutations, namely permutations that contain cycles of length not larger than 2. Said in different terms, they may only contain disjoint transpositions, that is permutations that swap two elements in a ``monogamic'' manner (an element can belong at most to one swapping couple).
If $\boldsymbol{\sigma}_{\mathcal M}$ is the permutation that swaps state $|a\rangle$ with $|b\rangle$, state $|c\rangle$ with $|d\rangle$ etc., then $U$ is a unitary that maximally mixes the same states and leaves all other states $|x\rangle$ unaltered, e.g.,
\begin{eqnarray}
U|a\rangle = \frac{|a\rangle + |b\rangle}{\sqrt{2}} \qquad
U|b\rangle = \frac{|a\rangle - |b\rangle}{\sqrt{2}} \\
U|c\rangle = \frac{|c\rangle + |d\rangle}{\sqrt{2}}\qquad
U|d\rangle = \frac{|c\rangle - |d\rangle}{\sqrt{2}} \\
\vdots\\
U|x\rangle = |x\rangle, \qquad x\neq a,b,c,d, \dots
\end{eqnarray}
It follows that the metrotropy is achieved by a unitary basis change that involves only maximally mixing partial swaps between distinct ``monogamic'' couples of states. This concludes our search for the minimum of $E'$ over unistochastic matrices.

Using the definition of metrotropy, $\mathcal{M} =E-  \min E' = E - (v_0+E)/2$, we finally obtain:
\begin{eqnarray}
\mathcal{M} = (E-v_0)/2
\end{eqnarray}

\subsection{Remarks}
If $\boldsymbol{\sigma}_\mathcal{W}$ is symmetric, then  the smallest among the $v_{ij}$'s coincides in value with $u_0$. Accordingly we have 
\begin{eqnarray}
\mathcal{M}=\mathcal{W}/2\\
 \boldsymbol{\sigma}_\mathcal{W}=\boldsymbol{\sigma}_\mathcal{M} 
\end{eqnarray}
It is not difficult to see that,
\begin{eqnarray}
\mathcal{M} \leq \mathcal{W}/2
\label{M<=W/2}
\end{eqnarray}
from which it follows that $\mathcal{W}=0 \Rightarrow \mathcal{M}=0$.

In general, when the initial state of the system is not stationary ($[H,\rho]\neq 0$), the average final energy after a unitary evolution reads:
\begin{eqnarray}
E' = \mathbf{E}^T \cdot\mathbf{Q} \cdot \mathbf{r}
\label{eq:E'coomuting}
\end{eqnarray}
where $Q_{nk}=|\langle n | U| r_k \rangle |^2$ and $|r_k\rangle$ denotes the eigenvectors of the state density matrix, $\rho$, namely $\rho=\sum r_n |r_n\rangle \langle r_n|$. The minimum is reached when $U$ maps the eigenstate of $H$ with highest eigenvalue, to  the eigenstate of $\rho$ with lowest eigenvalue; the eigenstate of $H$ with second highest eigenvalue, with the eigenstate of $\rho$ with second lowest eigenvalue, etc... . This results, just like in the commuting case, into $\mathbf{Q}$ being a permutation, and the according final energy  $E'$ coinciding with the minimum that can be reached by any bistochastic matrix.

The average final energy after a projective measurement reads in the case $[H,\rho]\neq 0$:
\begin{eqnarray}
E' = \mathbf{E}^T \cdot\mathbf{P}^T \cdot\mathbf{S} \cdot \mathbf{r}
\label{eq:E'coomuting}
\end{eqnarray}
where $P_{nk}= |\langle n | \psi_k \rangle |^2$ and $S=|\langle r_n | \psi_k \rangle |^2$. Since both $\mathbf{P}$ and $\mathbf{S}$
are bistochastic, so is their product $\mathbf{P}^T \cdot\mathbf{S}$. It follows that the minimum of $\mathbf{E}^T \cdot\mathbf{P}^T \cdot\mathbf{S} \cdot \mathbf{r}$ cannot be smaller than the minimum of $E' = \mathbf{E}^T \cdot\mathbf{Q} \cdot \mathbf{r}$. Accordingly, generally it is:
\begin{eqnarray}
\mathcal M \leq \mathcal W \, . 
\end{eqnarray}
Whether the relation $\mathcal M \leq \mathcal W/2$ holds in the case $[H,\rho]\neq 0$ remains an open question. Further studies are in order to assess the metrotropy in the case when the system is not initially in a stationary state, and/or one allows for higher rank projectors.

 The metrotropy naturally emerges in the study of measurement fuelled heat engines, for example two-stroke two-qubit engines \cite{Buffoni19PRL122}. For such engines the ergotropy is achieved by means of a swap operation on the two qubits, namely an overall symmetric permutation. Accordingly the metrotropy is achieved by means of a projection onto the singlet/triplet basis and is half the ergotropy \cite{Buffoni19PRL122}.

\section{Illustrative examples}
\subsection{3-level system}
\begin{figure}[t]%
    \centering
    \subfloat[
    ]{{\includegraphics[width=8cm]{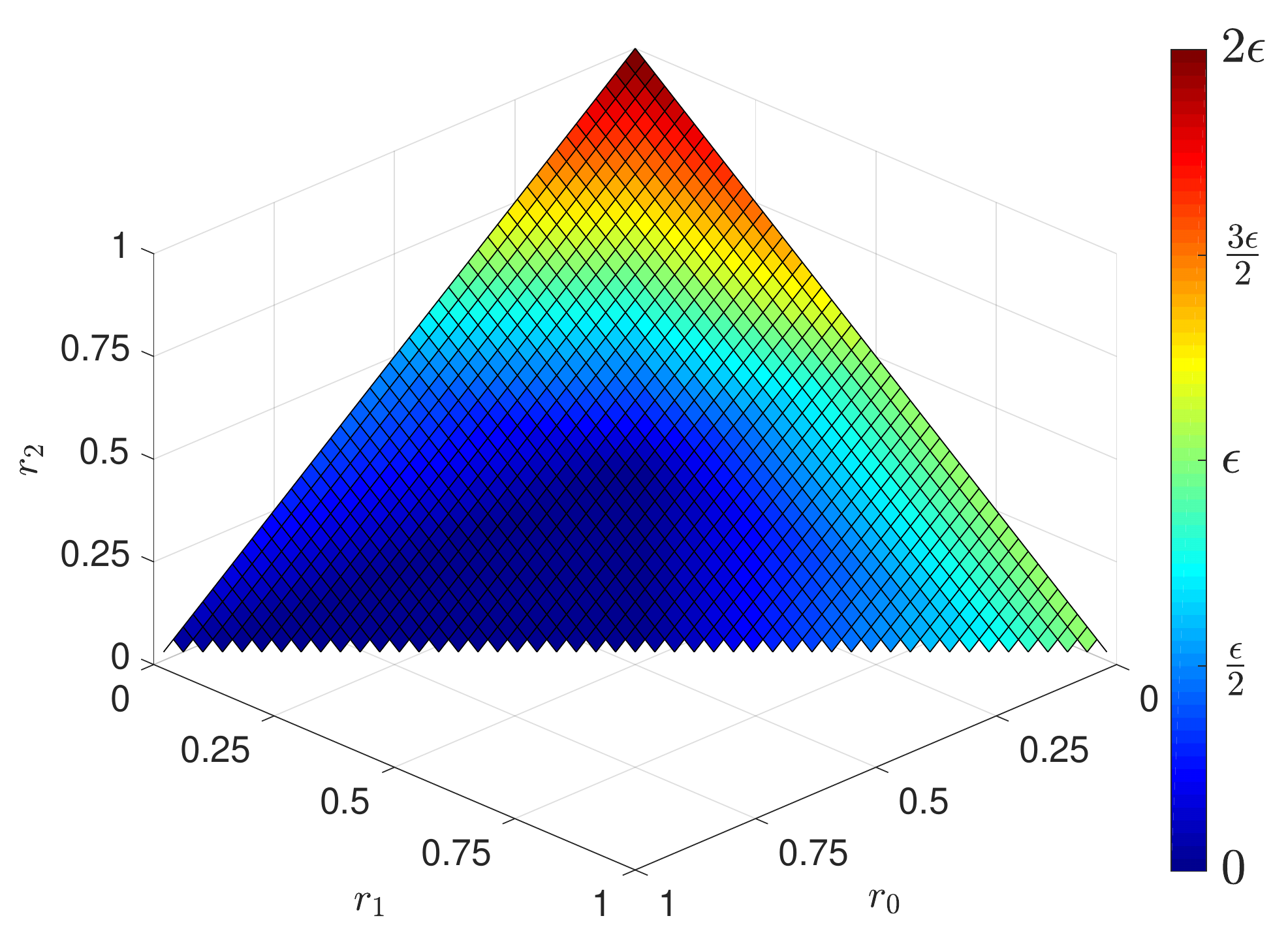} }}%
    \qquad
    \subfloat[
    ]{{\includegraphics[width=8cm]{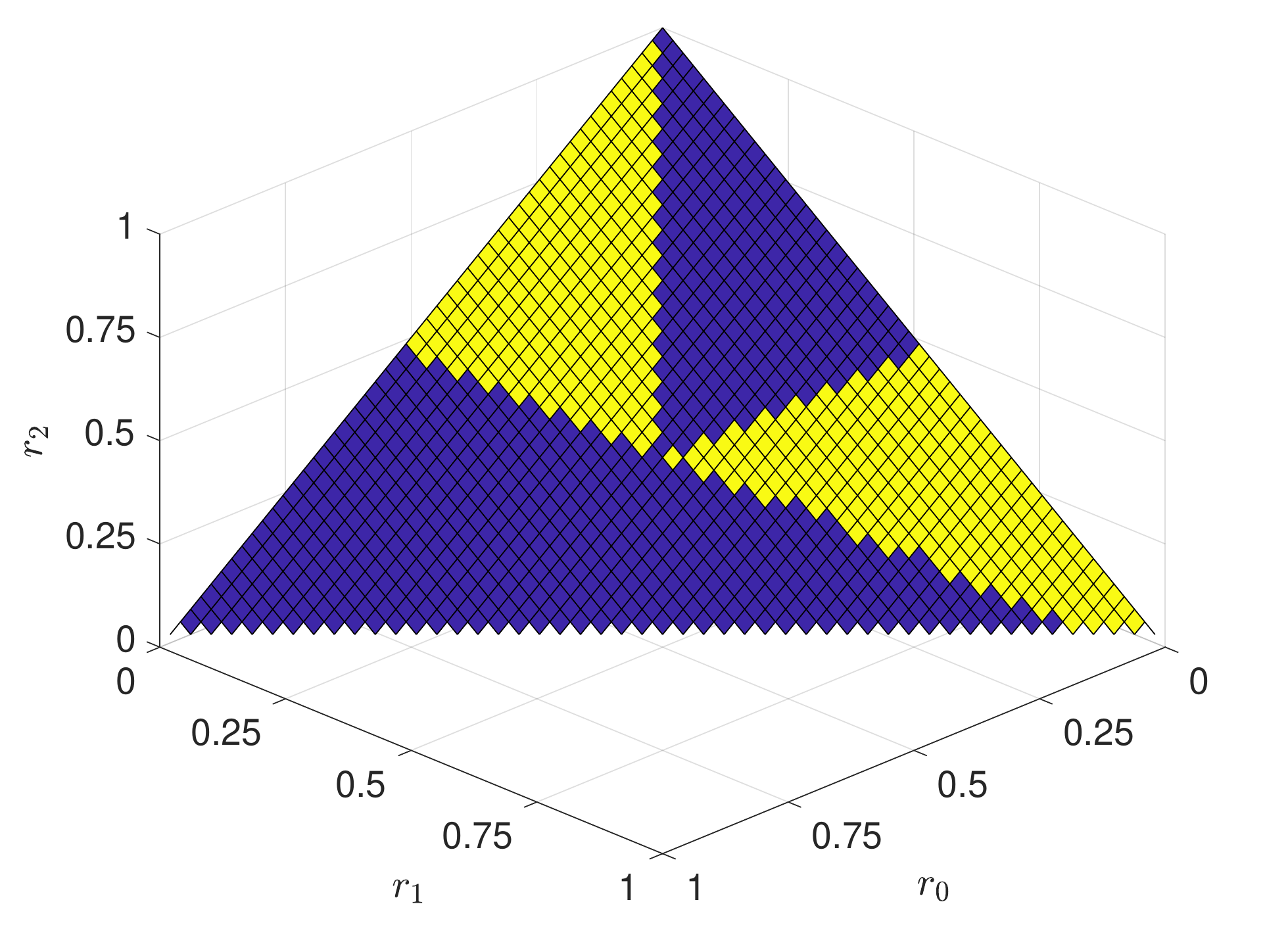} }}%
    \caption{Panel a): Ergotropy, $\mathcal{W}$. Panel b): Symmetricity of the ergotropy permutation $\mathbf{\sigma}_{\mathcal{W}}$; blue denotes symmetric $\mathbf{\sigma}_{\mathcal{W}}$ , yellow denotes non-symmetric $\mathbf{\sigma}_{\mathcal{W}}$.}
    \label{fig:ergotropy}%
\end{figure}

We consider a 3-level system with Hamiltonian: 
\begin{eqnarray}
H=\epsilon\left(\begin{array}{ccc}-1 & 0 & 0 \\0 & 0 & 0 \\0 & 0 & 1 \end{array}\right)
\end{eqnarray}
Figure \ref{fig:ergotropy}a shows the ergotropy $\mathcal{W}$ as a function of the populations of the most excited state ($r_2$), of the mid-energy state ($r_1$) and the lowest energy state ($r_0$). Note that the normalisation condition $r_0+r_1+r_2=1$, with $r_{0,1,2}\geq 0$ constraints the plot to the equilateral triangle with vertices $(0,0,1)^T,(0,1,0)^T,(1,0,0)^T$. Note that the ergotropy $\mathcal W$ is maximal when the system is initially in the  most excited state $(0,0,1)^T$ in which case it attains the value $2\epsilon$, it is null at $(1,0,0)^T$, and takes the value $\epsilon$ at $(0,1,0)^T$. Figure \ref{fig:ergotropy}b shows in blue the points where the ergotropy permutation $\mathbf{\sigma}_{\mathcal{W}}$ is symmetric and in yellow the points where it is not symmetric. The regions boundaries lie on the triangle edges and bisectrices.

\begin{figure}[t]%
    \centering
    \subfloat[
    ]{{\includegraphics[width=8cm]{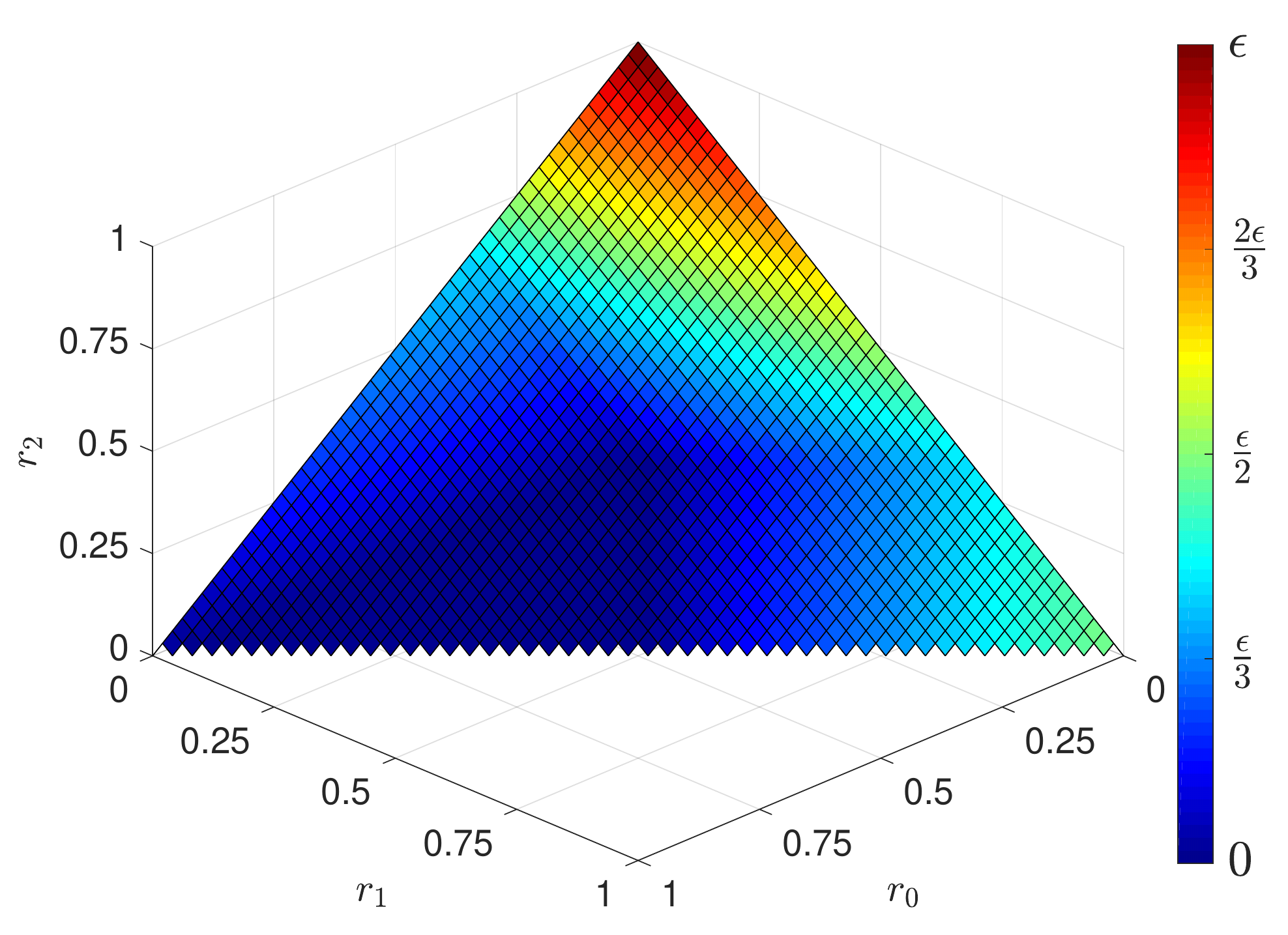} }}%
    \qquad
    \subfloat[
    ]{{\includegraphics[width=8cm]{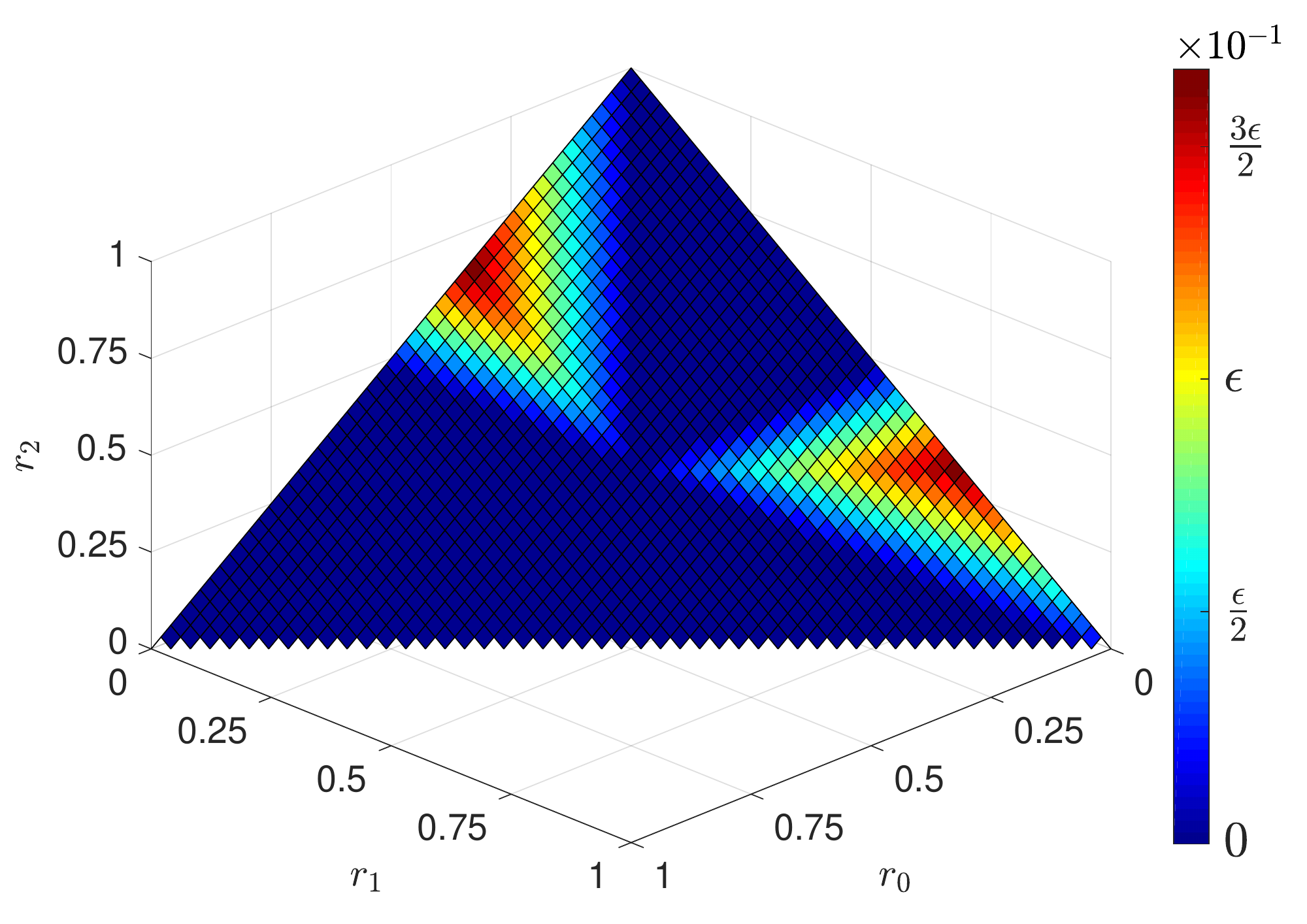} }}%
    \caption{Panel (a): Metrotropy $\mathcal{M}$. Panel b): $\mathcal{W}/2-\mathcal{M}$.}
    \label{fig:metrotropy}%
\end{figure}

Figure \ref{fig:metrotropy}a shows the metrotropy $\mathcal M$. The plotted data have been obtained by numerical minimisation of the final energy $E' = \mathbf{E}^T \cdot\mathbf{P}^T \cdot\mathbf{P} \cdot \mathbf{r}$ over all matrices $\mathbf{P}$ of the form $P_{kl}=|\langle k|U|l\rangle |^2$. The minimisation has been accordingly performed over all unitaries  belonging to $SU(3)$. To that end we have employed the parameterization of $SU(3)$ described in \cite{Byrd97arXiv}. As predicted by our theory the minimum of $E'$ was always attainted by unistochastic matrices of the form $\mathbf{P}=(\mathbb{1}+\boldsymbol{\sigma}_{\mathcal{M}})/2$ where $\boldsymbol{\sigma}_{\mathcal{M}}$ is a symmetric permutation. Note that the metrotropy is maximal at $(0,0,1)^T$ (attaining the value $\epsilon$), it is null at $(1,0,0)^T$, and takes the value $\epsilon/2$ at $(0,1,0)^T$ corresponding in all cases to half the according ergotropy.

Figure \ref{fig:metrotropy}b shows the difference $\mathcal{W}/2-\mathcal{M}$. Note that it is non-negative, it is null where the ergotropy permutation is symmetric and it is non-null elsewhere, in accordance to our predictions.

\subsection{2-level system in Bloch representation}
Let the Hamiltonian be $H=b_z \sigma_z$, and $\rho=(\mathbb{1}+\mathbf{R}\cdot \boldsymbol{\sigma})/2$ where $\boldsymbol{\sigma}=(\sigma_x,\sigma_y,\sigma_z)^T$ is a compact notation for the three Pauli matrices, $b_z>0$ and  $0\leq |\mathbf{R}|\leq 1$ (note that in what follows the symbol $\sigma$ denotes a Pauli matrix and not a permutation). In this case there are only 2!=2 permutations, namely the identity, $\mathbb{1}$, and the permutation that swaps 
the two eigenstates of $\sigma_z$. It is instructive to use the Bloch representation to gain insight on the relation between ergotropy and metrotropy, and as well on how to treat the case of a non-stationary initial state. 

The energy is represented by a vector along the $z$ direction with length $b_z$. If the state is stationary its Bloch vector is also along $z$ with length $|R_z|$ and direction given by the sign of $R_z$. When $R_z$ is negative the ergotropy is null and so is the metrotropy. Otherwise the ergotropy is achieved by means of a $\pi$ rotation around any axis lying in the $xy$ plane, which realises the swap permutation. 
Its value is $\mathcal W = 2R_z b_z$.

The effect of a measurement in the energy eigenbasis is to cancel the off diagonal elements of the density matrix, namely to project the vector $\mathbf{R}$ onto the $z$ direction: $\mathbf{R}=(R_x,R_y,R_z) \rightarrow \mathbf{R'} = (0,0,R_z)$. Accordingly, the effect of a measurement onto a generic basis, is that of projecting the vector $\mathbf{R}$ onto a generic direction, represented by a unit vector $\mathbf{n}$ pointing somewhere on the Bloch sphere:
\begin{equation}
\rho'= \frac{\mathbb{1}+(\mathbf{R}\cdot \mathbf{n}) {\mathbf{n} \cdot \boldsymbol{\sigma}}}{2} = \frac{\mathbb{1}+\mathbf{R}'\cdot \boldsymbol{\sigma}}{2}
\end{equation}
One then sees that when $\mathbf{R}=(0,0,R_z)$ the metrotropy is achieved by projecting onto a direction perpendicular to the $z$ direction. This results in annihilating $\mathbf{R}=(0,0,R_z)\rightarrow (0,0,0)$, accordingly the metrotropy is $\mathcal M= R_z b_z$, that is half the ergotropy, as expected on the basis that the corresponding (swap) permutation is an involution. The post measurement density matrix would so result in the identity matrix.

\begin{figure}[t]%
    \centering
    \includegraphics[width=8cm]{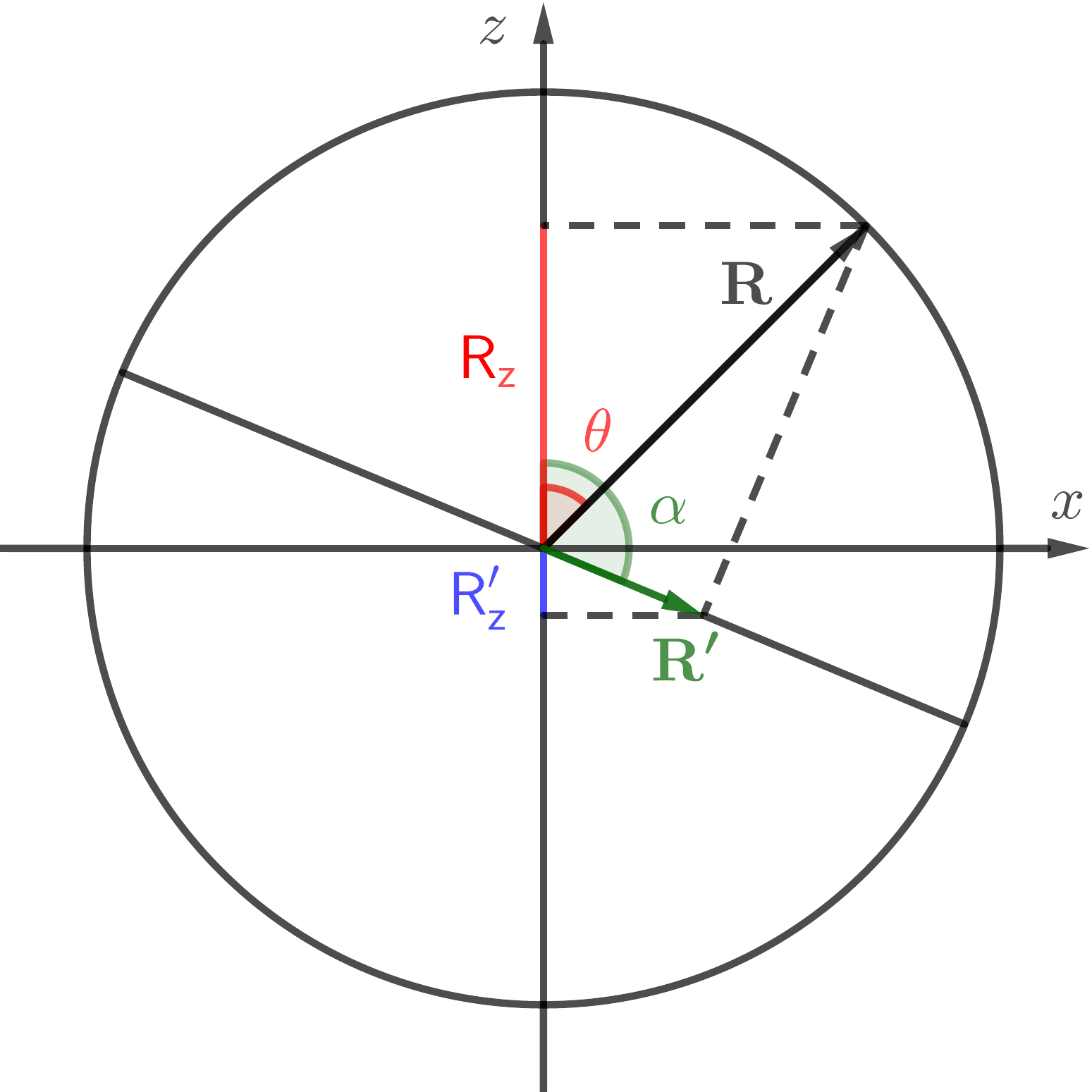} 
    \caption{As a consequence of a  generic projective measurement, the Bloch vector, $\mathbf{R}$, representing the state of a qubit, gets projected along some direction, and turns into a shorter vector $\mathbf{R}'$. Apart from a coefficient given by the intensity of the magnetic field $b_z$, the metrotropy is given by the difference of the $z$-component of the pre- and post- measurement Bloch vectors, ($\mathbf{R}$ and $\mathbf{R}'$, respectively), namely the length of the red-blue segment. }
    \label{fig:bloch}%
\end{figure}

When the initial state is not stationary, namely the Bloch vector $\mathbf{R}$ points in a generic direction, the ergotropy is achieved by rotating it so as to align it in the negative $z$ direction. Its value would be accordingly, $\mathcal {W}=b_z |\mathbf{R}| (1+ \cos\theta)$, where $\theta$ denotes the polar angle of $\mathbf{R}$. We first notice, based on a symmetry argument, that the metrotropy is achieved when projecting along a direction $\mathbf{n}= \mathbf{R'}/| \mathbf{R'}|$ that lies in the plane containing $\mathbf{R}$ and the $z$ axis (in Figure \ref{fig:bloch} we chose our reference frame in such a way that the latter coincides with the $xz$ plane). By calling $\alpha$ the polar angle of $\mathbf{n}$ and looking for the minimum of $R'_z$, namely the $z$ component of the Bloch vector $\mathbf{R}'$ representing the post measurement state, we see that the metrotropy is achieved for $\alpha=\pi/2+\theta/2$. The according metrotropy is $\mathcal M = b_z |\mathbf R| (1+ \cos\theta)/2$. That is, in the two-level system case the relation $\mathcal M = \mathcal W/2$ holds regardless of whether the system is initially in a stationary state. We also note that maximal ergotropy and metrotropy are achieved when $\theta=0$, i.e., the two-level system is initially in a stationary state, and among all stationary states, the case when  $|\mathbf R|=R_z=1$, (namely the system is  in the eigenstate of $H$ with highest energy), is the one with largest ergotropy and metrotropy, as expected.

\section*{References}
\bibliographystyle{iopart-num}

\providecommand{\newblock}{}

\end{document}